\newif\iftemplate
\templatefalse

\documentclass[a4paper,11pt]{article}
\pdfoutput=1 

\usepackage{jinstpub} 

\usepackage{lineno}

\title{\boldmath Characterization of irradiated RD53A pixel modules with passive CMOS sensors}

\author[a,1]{A. Jofrehei,\note{Corresponding author.}}
\author[b]{M. Backhaus,}
\author[a]{P. Baertschi,}
\author[a]{F. Canelli,}
\author[b]{F. Glessgen,}
\author[a]{W. Jin,}
\author[a]{\\B. Kilminster,}
\author[a]{A. Macchiolo,}
\author[a]{A. Reimers,}
\author[b]{B. Ristic,}
\author[b]{R. Wallny}
\author[]{on behalf of the CMS Tracker Group}

\affiliation[a]{University of Zurich (UZH),\\Zurich, Switzerland}
\affiliation[b]{ETH Zurich,\\Zurich, Switzerland}

\emailAdd{arash.jofrehei@cern.ch}

\abstract{We are investigating the feasibility of using CMOS foundries to fabricate silicon detectors, both for pixels and for large-area strip sensors. The availability of multi-layer routing will provide the freedom to optimize the sensor geometry and the performance, with biasing structures in poly-silicon layers and MIM-capacitors allowing for AC coupling. A prototyping production of strip test-structures and RD53A compatible pixel sensors was recently completed at LFoundry in a 150$\,$nm CMOS process. This paper will focus on the characterization of irradiated and non-irradiated pixel modules, composed by a CMOS passive sensor interconnected to a RD53A chip. The sensors are designed with a pixel cell of $25\times100\,\mu \mathrm{m}^2$ in case of DC coupled devices and $50\times50\,\mu \mathrm{m}^2$ for the AC coupled ones. Their performance in terms of charge collection, position resolution, and hit efficiency was studied with measurements performed in the laboratory and with beam tests. The RD53A modules with LFoundry silicon sensors were irradiated to fluences up to $1.0\times10^{16}\,\frac{\mathrm{n}_\mathrm{eq}}{\mathrm{cm}^2}$.}


\proceeding{The 12$^{\text{th}}$ International Conference on Position Sensitive Detectors\\
  12$^{\text{th}}$-17$^{\text{th}}$ of September 2021\\
  Birmingham}

\begin{document}
\maketitle
\flushbottom

\section{Introduction}
\label{sec:intro}
The LHC accelerator will be upgraded to deliver instantaneous peak luminosities of $5-7.5 \times 10^{34}\,\mathrm{cm}^{-2}\mathrm{s}^{-1}$ with 140-200 collisions per bunch crossing. This will allow ATLAS and CMS to collect integrated luminosities of the order of 300$\,\mathrm{fb}^{-1}$ per year and up to 3000$\,\mathrm{fb}^{-1}$ during
the HL-LHC projected lifetime of ten years. Under the hypothesis of a replacement after the end of Run 5, the CMS tracker will have to endure a fluence of $2\times10^{16}\,\frac{\mathrm{n}_\mathrm{eq}}{\mathrm{cm}^2}$ for its first layer~\cite{CMS}. The CMS Inner Tracker (IT) will undergo a complete upgrade, known as the Phase-2 upgrade, in which the increased granularity enhances the track and vertex reconstruction, and copes with the increased trigger rate. In this regard, pixel sensor prototypes are developed in collaboration with LFoundry~\cite{LF} using the CMOS technology which is widely used in the semiconductor industry. The large-scale production and the possibility of small on-pixel structures makes these sensors cheaper, faster to produce, and possibly with better performance compared to standard productions in high energy physics (HEP). The samples have a 150$\,\mu\mathrm{m}$ thickness with a pixel size of 25$\times$100$\,\mu\mathrm{m}^2$ and 50$\times$50$\,\mu\mathrm{m}^2$, are irradiated with 23$\,$MeV protons at the Irradiation Center in Karlsruhe~\cite{KIT}, and are either DC or AC coupled to the RD53A readout chip (ROC). The RD53A ROC is a prototype for the development of the final readout chip for the Phase-2 upgrade of the CMS IT and ATLAS pixel detector~\cite{RD53}. It consists of 192 rows and 400 columns with a pitch of $50\times50\,\mu\mathrm{m}^2$. The measurements were done using the DESY test beam infrastructures~\cite{DESY}. The measurements on the non-irradiated CMOS sensors can be found in Ref.~\cite{non-irradiated}.

\section{LFoundry CMOS technology}
\label{sec:CMOS}
CMOS technology is widely used in the semiconductor industry. Compared to the current technologies used in HEP detectors, it offers the possibility of small on-pixel structures. Features like 4 to 6 metal layers for signal redistribution, low- and high-resistivity polysilicon layers, and the possibility of mitigating leakage current noise with AC-coupled sensors improve the performance of silicon sensors. The samples in this study are passive planar n-in-p sensors for hybrid detector modules built in CMOS technology using the 150$\,$nm production line of LFoundry. The wafers are produced by implementing structures on reticule building blocks and applying the reticules on the photoresist. The reticules are then stitched to each other to build up the layout on 8-inch wafers. As most production costs are per-wafer, this will reduce the total production cost compared to the standard 6-inch wafers used for tracking detectors in HEP.

\section{Test beam results of irradiated RD53A modules}
\label{sec:results}
Four RD53A modules with LFoundry CMOS sensors listed in Table~\ref{tab:modules} were irradiated with 23$\,$MeV protons at the Irradiation Center in Karlsruhe~\cite{KIT}. Three modules were DC coupled and had a pixel pitch of 25$\times$100$\,\mu\mathrm{m}^2$, while one module was AC coupled and had a pixel pitch of 50$\times$50$\,\mu\mathrm{m}^2$. The devices were tested with a 5.2 GeV electron beam at the DESY test beam facility. The tracks traversing the device under test (DUT) are reconstructed using an AIDA telescope~\cite{AIDA}.
Irradiated RD53A modules are installed in a cooling box at $-35\,^\circ$C which can be precisely positioned and oriented. The RD53A data are read out with BDAQ53~\cite{BDAQ} and the event reconstruction is performed with the EUDAQ data acquisition framework~\cite{EUDAQ}.

\begin{table}[htbp]
\centering
\caption{\label{tab:modules} List of irradiated RD53A modules with CMOS sensors.}
\smallskip
\begin{tabular}{|c|c|c|c|}
\hline
Pitch ($\mu m^2$) & Type & Fluence ($\times 10^{15}\,\frac{\mathrm{n}_\mathrm{eq}}{\mathrm{cm}^2}$) & Threshold (e)\\
\hline
25 $\times$ 100 & DC & 10 & 1192\\
25 $\times$ 100 & DC & 2.1 & 1240\\
25 $\times$ 100 & DC & 9.2 & 1219\\
50 $\times$ 50 & AC & 5 & 1208\\
\hline
\end{tabular}
\end{table}

\subsection{Cluster hit efficiency}
\label{sec:efficiency}
\begin{figure}[htbp]
\centering 
\includegraphics[width=.47\textwidth]{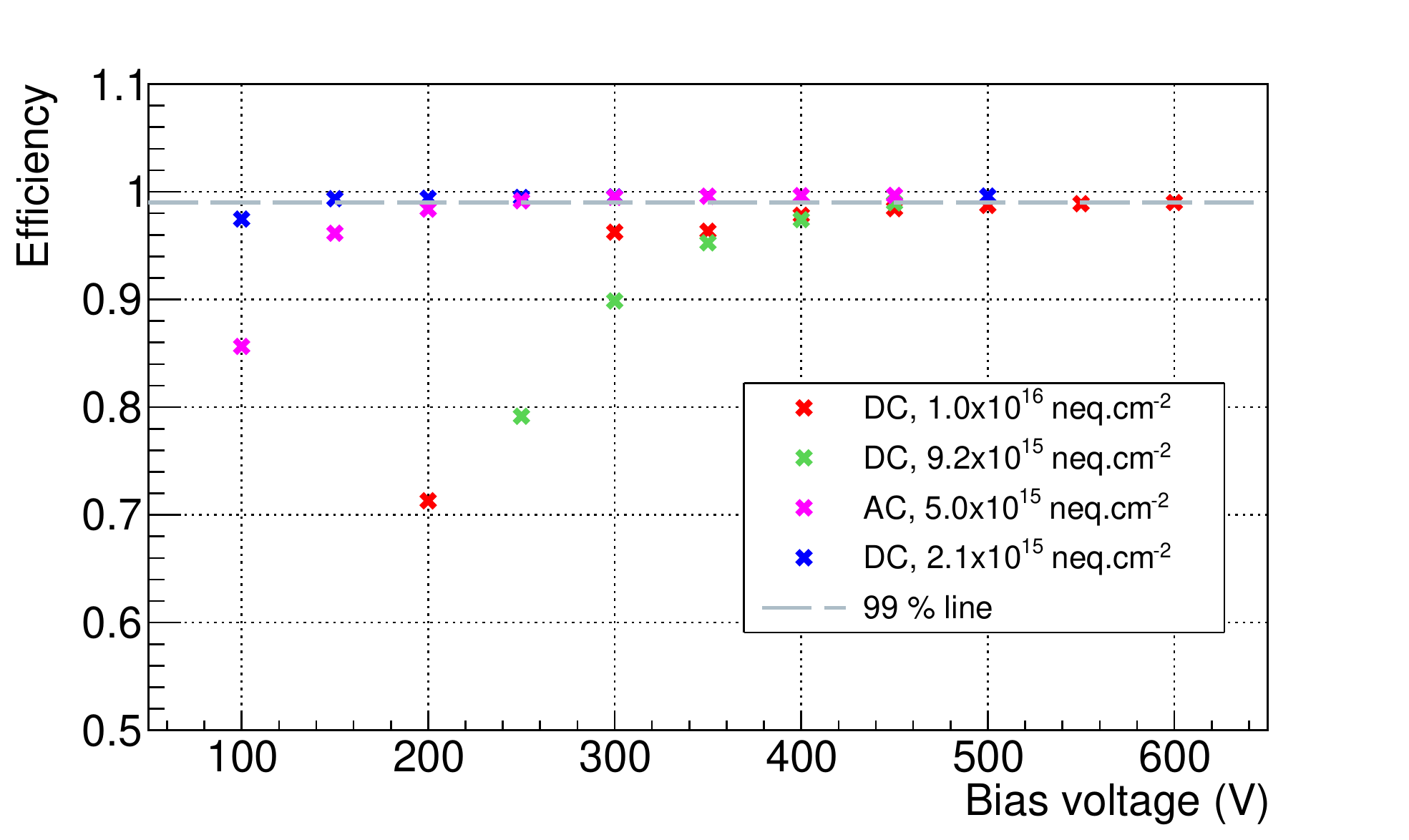}
\qquad
\includegraphics[width=.47\textwidth,]{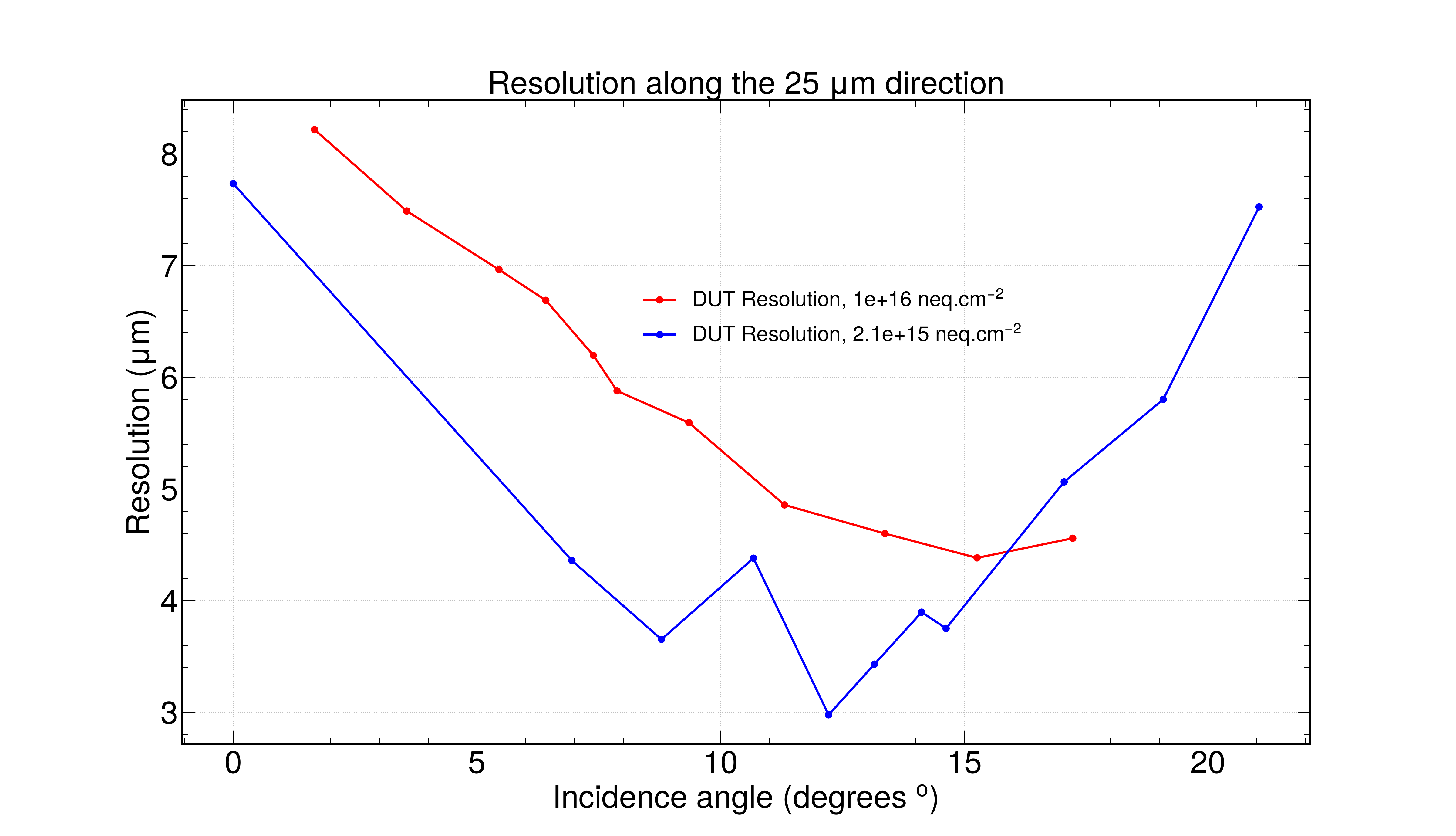}
\caption{\label{fig:efficiency_res} Hit efficiency (left) and position resolution (right) of irradiated RD53A modules with CMOS sensors.}
\end{figure}

The hit efficiency is defined as the ratio of the hits in the detector over the number of tracks passing through its active area. The efficiency increases with the bias voltage, as shown in Figure~\ref{fig:efficiency_res} (left). The probability of collecting charges above the readout threshold and therefore the efficiency decreases with irradiation, as also discussed in Section~\ref{sec:charge}. The required efficiency of 99\% is achieved for all the irradiated modules.

\subsection{Position resolution}
\label{sec:resolution}

The position resolution of a pixel detector is defined as the standard deviation of the difference between its reconstructed position with respect to the true position of the passing track. The measured position resolution is shown in Figure~\ref{fig:efficiency_res} (right) for two of the irradiated detectors in Table~\ref{tab:modules}. The distribution of the charge in the clusters and therefore the precision on the position reconstruction depends on the incident angle. For single-pixel clusters, which are dominant for perpendicular tracks, the position resolution is merely determined by the pixel pitch, while for double-pixel clusters the ratio of charge in the two pixels leads to a more precise position measurement. At high incident angles the uncertainty on the position starts to increase due to the digitization of relatively low pixel charges.

\subsection{Charge collection measurements}
\label{sec:charge}
The readout chip represents the signal amplitude with a digitized measurement of the time over threshold (TOT). Using the injection circuit of the RD53A chip, one can inject arbitrary amounts of charge into individual readout pixels. Besides other advantages for pixel-to-pixel calibrations and measurements, the charge injection is used to construct a mapping of TOT as a function of injected charge. The charge is given in the $\Delta$VCAL unit, a readout chip variable corresponding to about 10 electrons. This map can then be reversed so that for each given TOT of each individual pixel the corresponding charge can be evaluated. This provides a pixel-to-pixel calibration of cluster charges of test beam events.

\begin{figure}[htbp]
\centering 
\includegraphics[width=.29\textwidth]{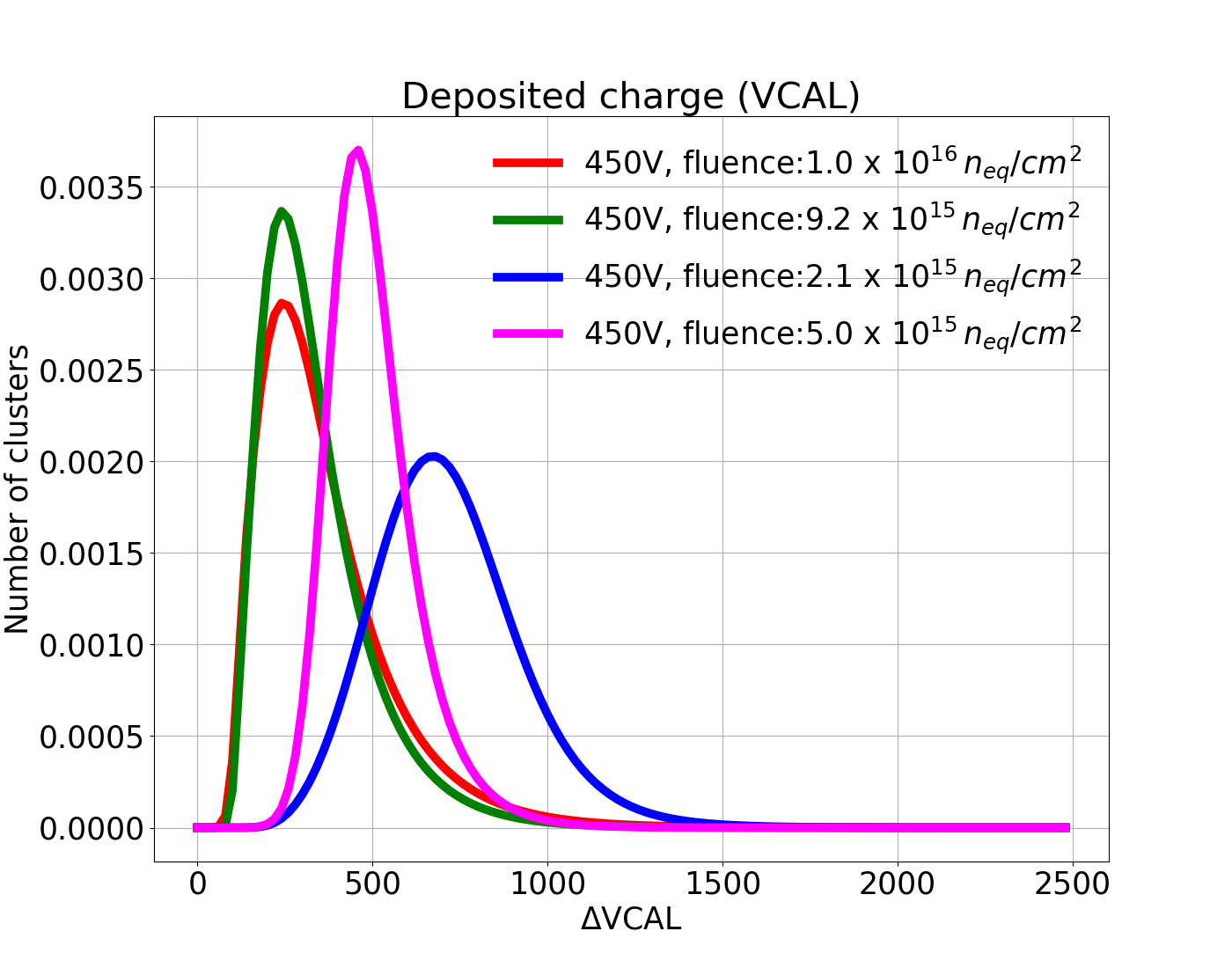}
\qquad
\includegraphics[width=.29\textwidth,]{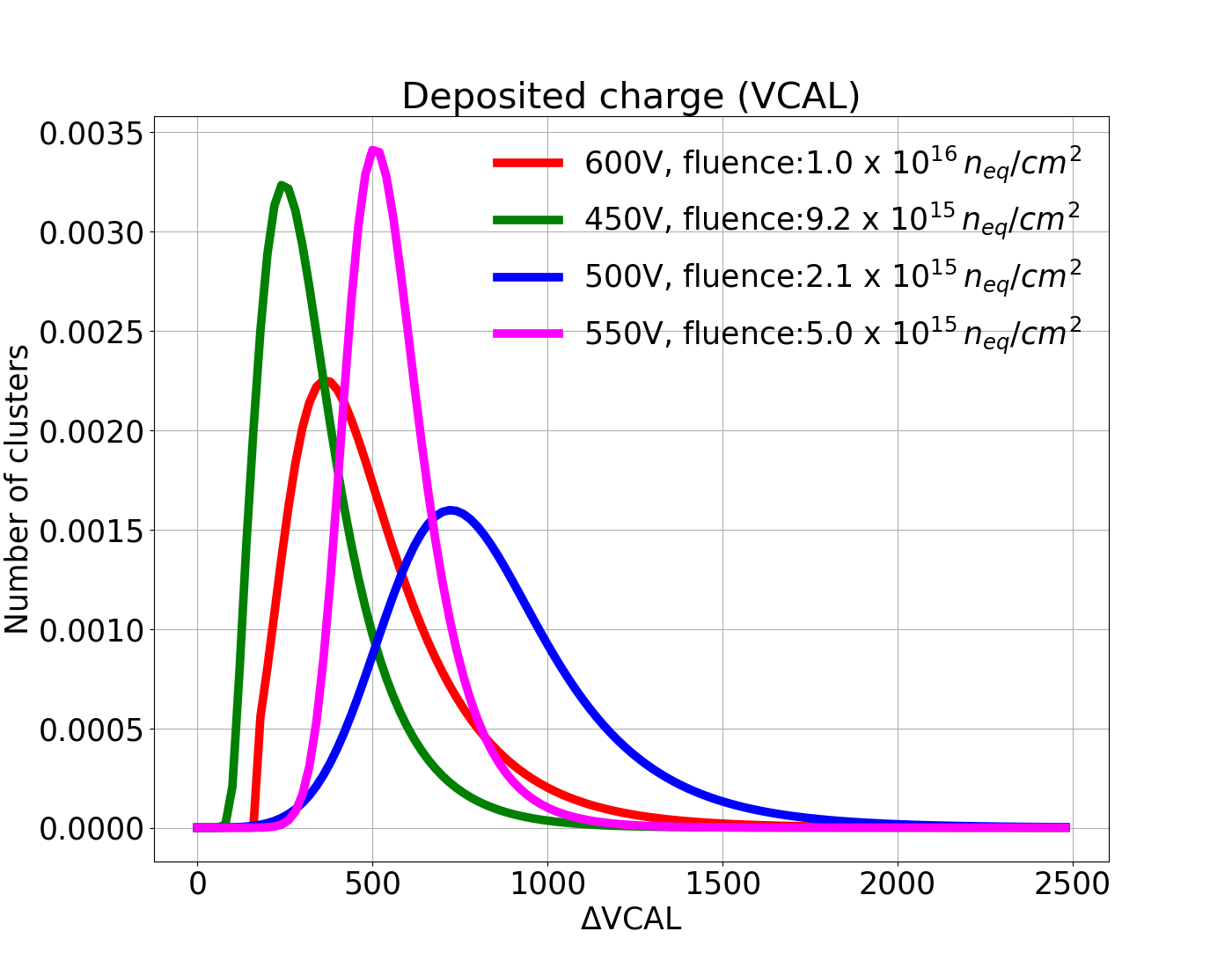}
\qquad
\includegraphics[width=.3\textwidth]{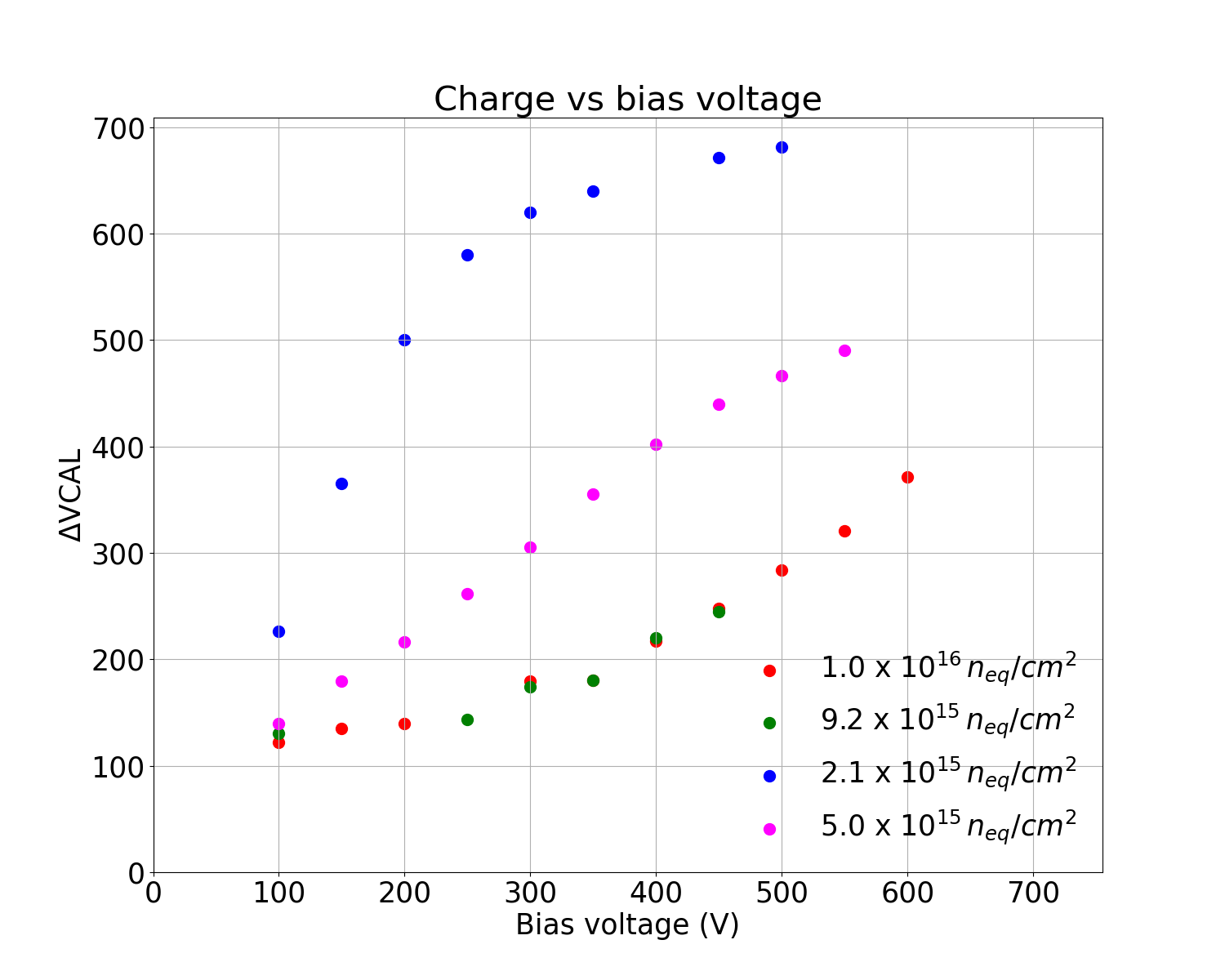}
\caption{\label{fig:chargeDist_mean} Distribution of the collected charge for the four irradiated RD53A modules with LFoundry sensors at a fixed bias voltage of 450V (left) and at the maximum bias voltage the modules reached in the test beam (center). The mean collected charge at each bias voltage decreases with fluence (right).}
\end{figure}

Figure ~\ref{fig:chargeDist_mean} depicts the cluster charge distributions for the four irradiated LFoundry modules. Increasing the fluence, the probability of charge trapping is higher and the depleted volume shrinks. Therefore, for a fixed applied voltage, the collected charge decreases with fluence.  The charge deposition in the silicon bulk can be described by a Landau distribution convoluted with the Gaussian contribution of the noise. Figure~\ref{fig:chargeDist_mean} shows the dependence of the collected charge on the fluence and the applied bias voltage. Less irradiated modules generally have a higher collected charge that saturates at lower bias voltages.

\subsection{High incident angles}
\label{sec:highEta}
The incident angle of charged particles in the first layer of the upgraded CMS pixel detector will be up to 75--85 degrees in high pseudorapidity ($\eta$) regions. Charged particles with high incident angles deposit charge in long clusters of pixels. For pixels of 25$\times$100$\,\mu\mathrm{m}^2$ size, the tracks at high $\eta$ release charge along the 100$\,\mu$m side, with an effective path in silicon smaller than for the perpendicular tracks. This could lead to broken clusters, which are challenging to reconstruct. We have studied the cluster breaking rate for the non-irradiated LFoundry modules using the test beam data taken at high incident angles.

\begin{figure}[htbp]
\centering 
\includegraphics[width=.3\textwidth]{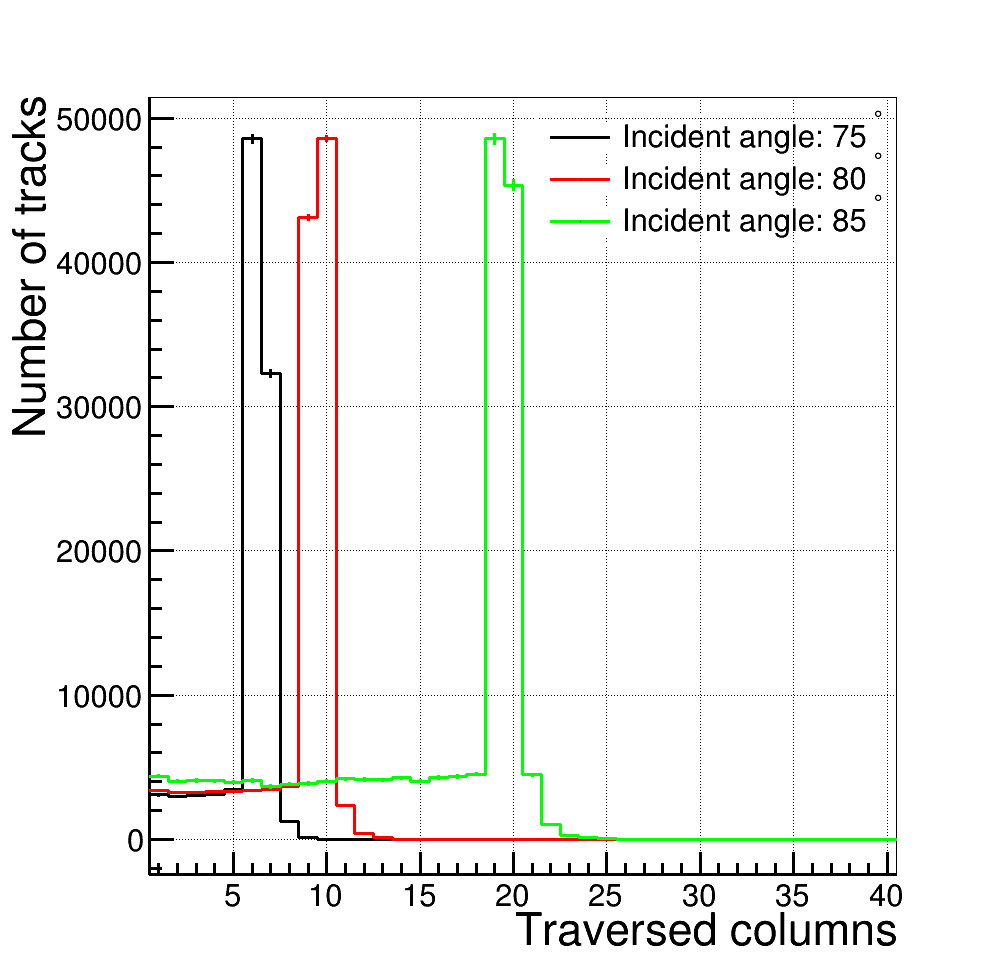}
\qquad
\includegraphics[width=.3\textwidth,]{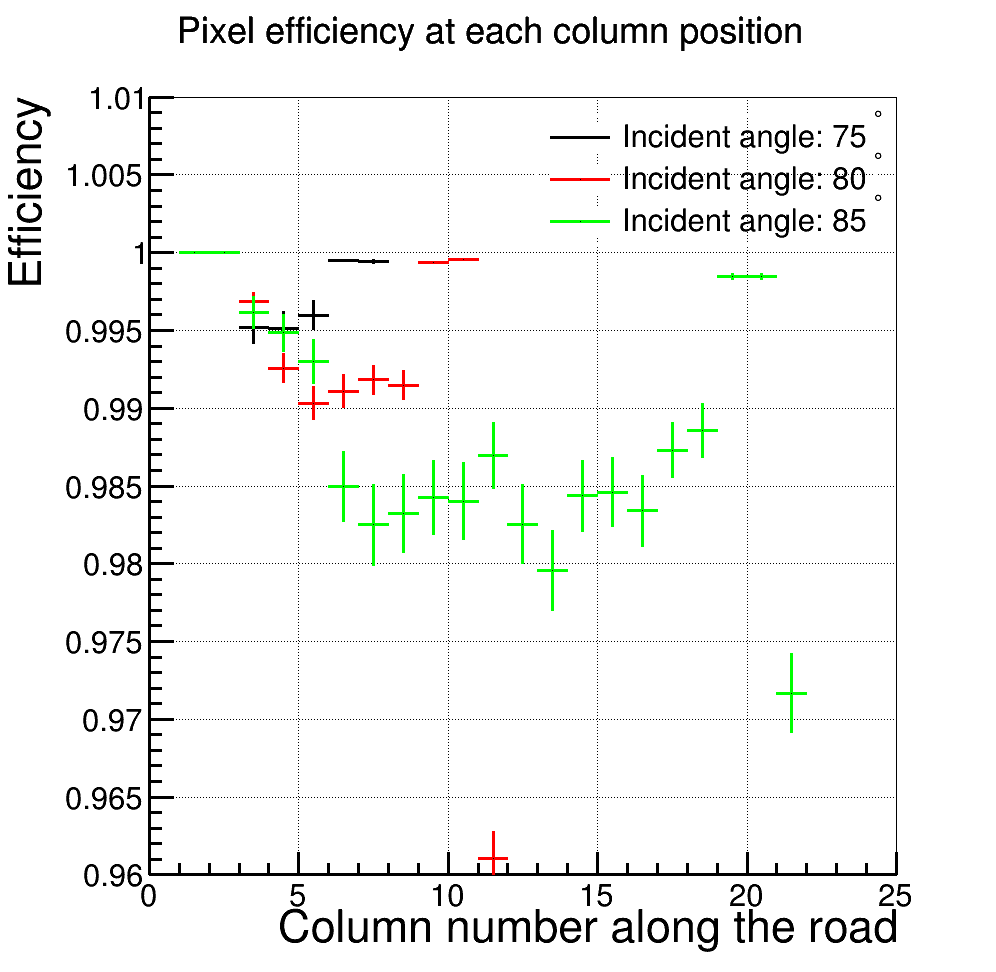}
\caption{\label{fig:highEta} Expected length of clusters for tracks at high incident angle (left) and the pixel efficiency of reconstructing a pixel along the track road at each incident angle (right).}
\end{figure}

After aligning the telescope tracks with pixel hits, a road can be defined for each track that passes through the silicon detector, which consists of the pixels along the track path. The length of a road is represented by the number of pixel columns it contains, while each column has a width of 100$\,\mu$m. Figure~\ref{fig:highEta} shows the distribution of the path length, expressed in number of traversed columns, for the reconstructed tracks at different incident angles. The distributions peak at the geometrically expected number of pixels, with a left tail corresponding to the tracks that were not reconstructed at the desired angles, due to multiple scattering and other uncertainties on the track incident angle. Keeping only the clean tracks from the peak of this distribution, one can loop through the pixels along this road that are expected to have a hit, and measure the reconstruction efficiency at each column position as depicted in Figure~\ref{fig:highEta}, right. The high reconstruction efficiency is promising, but the feasibility of tracking at high $\eta$ has to be validated by measurements with irradiated modules.

\section{Conclusions}
\label{sec:conclusion}
Irradiated RD53A modules with passive CMOS sensors are studied. The CMOS technology offers small on-pixel structures which can improve the performance of the pixel detectors by implementing AC coupling  or  using the intermediate metal layers for rerouting the signals between pixel cells and read-out chip channels. The use  of the stitching method to produce large wafers reduces the production cost, even if the initial set-up can be expensive. Measurements in the laboratory and test beams show comparable efficiency and resolution compared to standard productions~\cite{standard}, meeting the Phase-2 requirements for the CMS pixel detector.

\acknowledgments

We are greatly thankful to our colleagues of the University of Bonn, Y. Dieter, J. Dingfelder, \\T. Hemperek, F. Huegging, D. Pohl, T. Wang, and N. Wermes, as well as D. Münstermann of the University of Lancaster, who carried out the design and production of these LFoundry samples.


\end{document}